\documentclass[reprint,superscriptaddress,
aps,pra,
amsmath,amssymb]{revtex4-1}
\usepackage{graphicx}
\usepackage{dcolumn}
\usepackage{bm}
\usepackage{color}
\usepackage{natbib}
\usepackage{graphicx}
\usepackage{hyperref}
\usepackage{textgreek}

\usepackage{titlesec} 
\titlespacing\section{0pt}{10pt}{4pt}
\titlespacing\subsection{0pt}{10pt}{2pt}
\titlespacing\subsubsection{0pt}{10pt}{1pt}

\newcommand{\thetaF}{\theta_{_{\text{F}}}}

\begin{document}
\title{Rapid parameter determination of discrete damped sinusoidal oscillations}
\author{Jim C. Visschers}
\affiliation{Institut f\"ur Physik, Johannes Gutenberg Universit\"at-Mainz, 55128 Mainz, Germany}
\author{Emma Wilson}
\affiliation{Department of Mathematics, University of Leicester, Leicester, LE1 7RH, UK}
\affiliation{Photek Ltd., St Leonards on Sea, East Sussex, TN38 9NS, UK}
\author{Thomas Conneely}
\affiliation{Photek Ltd., St Leonards on Sea, East Sussex, TN38 9NS, UK}
\author{Andrey Mudrov}
\affiliation{Department of Mathematics, University of Leicester, Leicester, LE1 7RH, UK}
\author{Lykourgos Bougas}\email{lybougas@uni-mainz.de}
\affiliation{Institut f\"ur Physik, Johannes Gutenberg Universit\"at-Mainz, 55128 Mainz, Germany}
\date{\today}

\begin{abstract}
We present different computational approaches for the rapid extraction of the signal parameters of discretely sampled damped sinusoidal signals. We compare time- and frequency-domain-based computational approaches in terms of their accuracy and precision and computational time required in estimating the frequencies of such signals, and observe a general trade-off between precision and speed. Our motivation is precise and rapid analysis of damped sinusoidal signals as these become relevant in view of the recent experimental developments in cavity-enhanced polarimetry and ellipsometry, where the relevant time scales and frequencies are typically within the $\sim1-10$\,\textmu s and $\sim1-100$\,MHz ranges, respectively. In such experimental efforts, single-shot analysis with high accuracy and precision becomes important when developing experiments that study dynamical effects and/or when developing portable instrumentations. Our results suggest that online, \textit{running}-fashion, microsecond-resolved analysis of polarimetric/ellipsometric measurements with fractional uncertainties at the $10^{-6}$ levels, is possible, and using a proof-of-principle experimental demonstration we show that using a frequency-based analysis approach we can monitor and analyze signals at kHz rates and accurately detect signal changes at microsecond time-scales.
\end{abstract}

\maketitle

\section{Introduction}
Precise and rapid signal-parameter estimation is important for both fundamental and applied research, and becomes particularly crucial when observing and controlling fast processes in real time (e.g., chemical reactions), and in the development of portable instrumentation where fast, real-time, data streaming and inspection is essential. \\
\indent Several different research fields rely on the precise and accurate extraction of the time constants and frequencies of damped sinusoidal signals. Prominent examples include: nuclear magnetic resonance (NMR)\,\cite{gunther2013nmr}, where information on the structure and the spin environment of a target molecule is extracted from precise determination of the frequency and decay constant of a damped sinusoidal signal; free-induction-decay (FID) optical magnetometry\,\cite{Savukov2005,gemmel2010ultra,Nikiel2014,Grujic2015,Hunter2018,Hunter2018OptExp}, where the magnetometric sensitivities depend on the precision of the measurement of the oscillating frequency; and pulsed/continuous-wave cavity ring-down polarimetry (CRDP)\,\cite{Mueller2000,Muller2002,sofikitis2014evanescent,bougas2015chiral,Dupre2015,SPILIOTIS2020,Spiliotis2020Laser,visschers2020continuous} and ellipsometry (CRDE)\,\cite{Papadakis2011,stamataki2013monitoring,Sofikitis2013,Sofikitis2015SDR}, where polarization-dependent absorption and refraction/reflection through/by an optical medium is extracted with high sensitivity through the precise measurement of the signal-decay time and its polarization beat frequency.\\
\indent A distinction among the aforementioned examples can be made according to their respective decay constants and oscillating frequencies. In routine NMR, typical decay times are in the $10^{-2}\!-\!10$\,s range, while frequencies are in the $10\!-\!800$\,MHz range; especially, portable NMR instruments operate in the $10-30$\,MHz frequency range\,\cite{Lee2008,Perlo1279,Lei2017,Lei2020}. In FID optical magnetometry, typical decay times are in the $10^{-2}-1$\,s range, while frequencies are within the $10^2-10^5$\,Hz range (see for instance, Refs.\,\cite{Savukov2005,Grujic2015}). In CRDP/CRDE demonstrations, however, decay times are typically in the $10^{-7}-10^{-5}$\,s range, while polarization beat frequencies are in the $1\!-\!100$\,MHz range. \\
\indent For all these applications, significant data processing is typically required to determine the signal parameters, and, in general, experimental sensitivity is improved by averaging over many measurement runs. As such, when developing portable instruments one needs to appropriately select the instrument's sampling and acquisition rates, but also carefully consider the computational cost, i.e. the calculation time, to analyze each acquired signal. For applications where the relevant time-scales are relatively long (10\,ms - 1\,s), such as NMR or FID magnetometry, there are several options that can provide precise results sufficiently fast (with respect to ``single-events"), such as, e.g., frequency counters (see Refs.\,\cite{PRIGL1996,Dong2016} and references therein). However, in applications where the relevant time-scales are much shorter than a few ms, as in the case of CRDP/CRDE, acquisition and computational speeds ultimately define, respectively, the measurement and analysis repetition rates. \\
\begin{figure*}[ht!]
\centering
    \includegraphics[width=\linewidth]{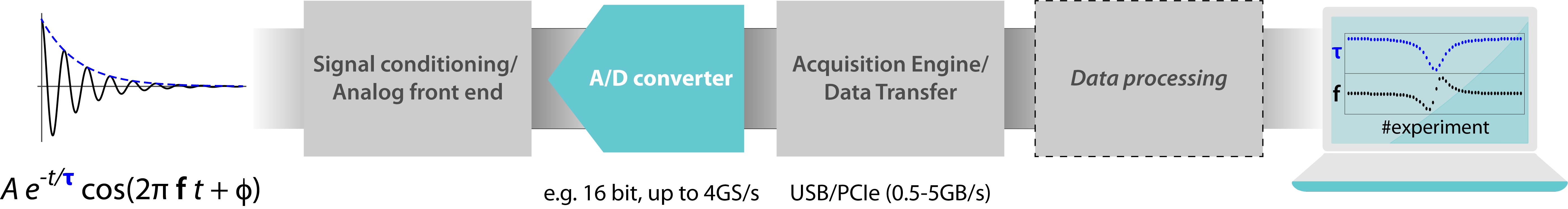}
    \caption{\small{\textit{Data acquisition block diagram -} a damped sinusoidal, physical, signal acquired by an experimental apparatus is digitized using an analog-to-digital (A/D) converter and then transferred to a data processing unit, which estimates the signal parameters and displays them on screen for real-time monitoring.}}
    \label{fig:Diagram}
\end{figure*}
\indent It is instructive to consider here a general data acquisition system, as depicted in Fig.\,\ref{fig:Diagram}: such a system collects the physical signal, in our case a damped sinusoidal signal, through an analog front-end that typically performs some signal conditioning (e.g. analog filtering, signal amplification). The acquired signal is subsequently digitized using an analog-to-digital converter and transferred to a data processing unit for parameter analysis. A particular case example  is the field-programmable gate array (FPGA), whose development has lead to the emergence of stand alone multi-channel (e.g., four) high-precision (e.g., 16-bit) data acquisition systems with high sampling (e.g., 2\,GS/s) and triggering rates (e.g., 1\,MHz), which have nowadays become commercially available at cost-effective rates. In such systems, data are (typically) transferred via USB or PCIe interfaces and, as such, data transfer rates as high as 5\,GB/s are feasible (e.g. Teledyne SP Devices, ADQ series\,\footnote{https://www.spdevices.com}). Using such systems, therefore, it is possible to perform (sub-)microsecond-resolved CRDP/CRDE measurements\,\cite{Sofikitis2015SDR}. However, the principal limiting factors towards an online, real-time, processing system that needs to be operable in a \textit{running} fashion with no dead-times, is the data-processing module of the overall data acquisition system (Fig.\,\ref{fig:Diagram}) and its limited memory storage capacity. Considering that in CRDP/CRDE experiments demonstrated decay time constants are in the $10^{-7}-10^{-5}$\,s range, it is important to identify appropriate computational approaches that can be implemented in data acquisition systems, such as FPGA-based digitizers, to allow for online, real-time, signal analysis at such fast time scales.\\
\indent Time- and frequency-based computational methodologies for rapid parameter estimation have been developed within the context of CRD spectroscopy, and these have been evaluated and compared in terms of their speed and precision\,\cite{Halmer2004,Mazurenka2005,everest2008discrete}. Most notably, Fourier transform methods have been implemented on FPGAs for fast analysis of exponentially decaying signals\,\cite{Spence2012}, with demonstrated analysis rates as high as 4.4\,kHz\,\cite{bostrom2015discrete}. However, while several works discuss the performance of various time- and frequency-domain analysis algorithms for damped sinusoidal signals\, \cite{aboutanios2011estimating,aboutanios2009estimation}, a direct comparison between their attainable precision and computational speed is currently missing. \\
\indent In this work we compare three specific analysis methods of discretely sampled damped sinusoidal signals in terms of their speed, and attainable accuracy and precision. These methods are: (a) a time-domain least-squares analysis based on a Levenberg-Marquardt algorithm\,\cite{more1978levenberg}; (b) a frequency-domain analysis based on a fast Fourier algorithm\,\cite{cooley1965algorithm,Mazurenka2005,Boyson2011,everest2008discrete} in combination with a quadratic interpolation of the frequency components of the resulting Fourier transform\,\cite{stamataki2013monitoring}; and (c) a time-domain analysis based on the Prony method\,\cite{wilson2019implementation}. We evaluate their efficacy in terms of the signal's parameters, and discuss how each of these affect the sensitivity limits for each computational methodology. Finally, we present an experimental, proof-of-principle, demonstration of the capabilities of such methods for the online analysis of CRDP signals.

\section{Theory}
\subsection{Damped sinusoidal signals}
\begin{figure}[ht]
\centering    \includegraphics[width=0.9\linewidth]{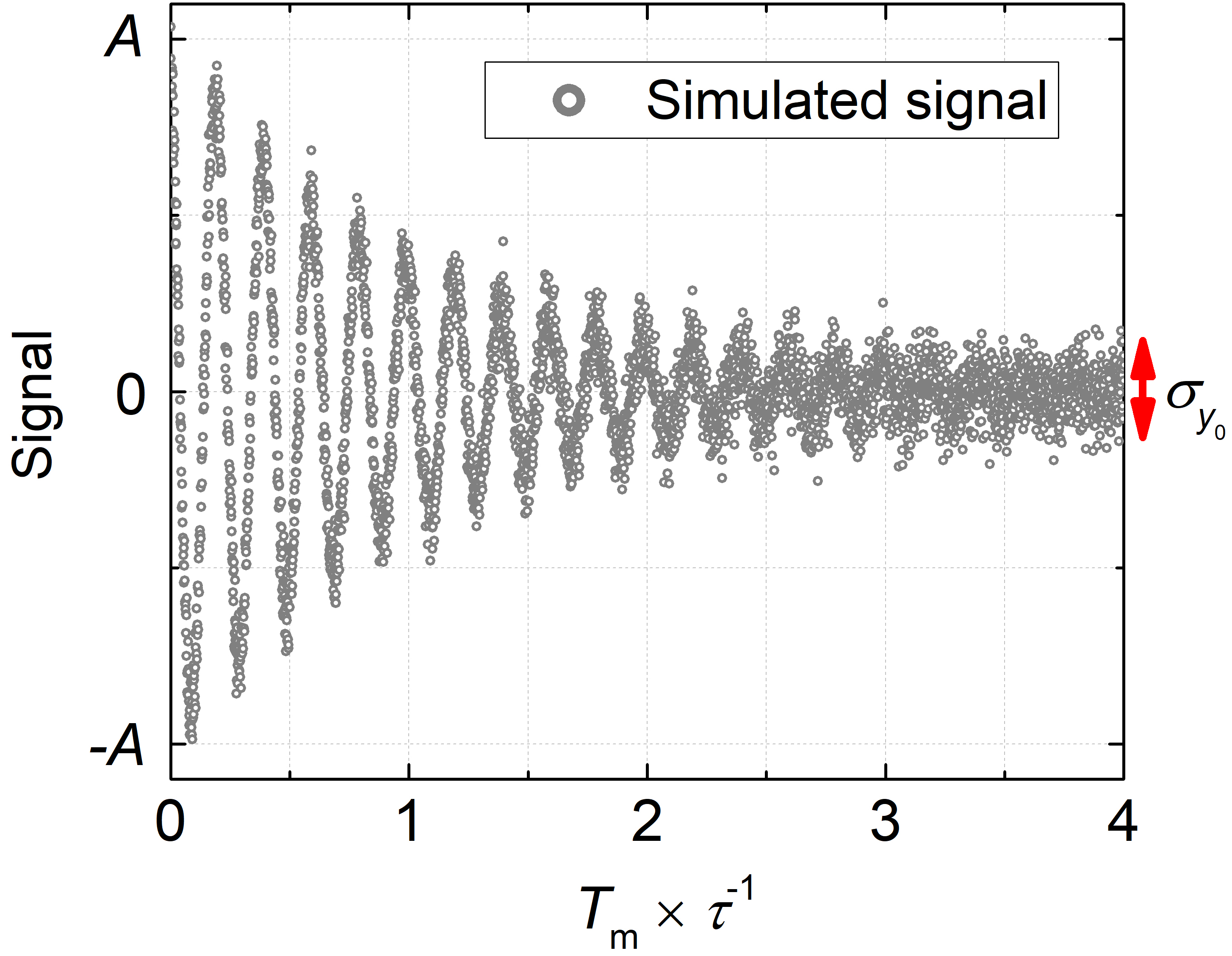}\vspace*{-3mm}
    \caption{\small{Example of a discretely sampled damped sinusoid as described by Eq.\,\ref{EQ: FID}, for $\text{f}\times\tau=5$, $n\times\tau^{-1}=200/\tau$, and SNR$\,=2^{4}$.}}
    \label{fig:Example Ringdown}
\end{figure}

\noindent A damped sinusoidal signal can be characterized in terms of a model function as:
\begin{equation}
    y(t)= A\cdot e^{-t/\tau} \cdot\text{cos}\left(2\pi\, \cdot{\rm{f}}\cdot t+\phi\right) + y_0,
\label{EQ: FID}
\end{equation}
where $t$ is the (discretely sampled) independent (time) variable of the signal, $A$ is the amplitude of oscillation, $\tau$ is the characteristic decay time, f and $\phi$ the frequency and phase of the oscillation, respectively, and $y_0$ is a global signal offset. Under realistic experimental conditions, all the signal parameters will be time-dependent, and the power spectral density of the signal will be proportional to their respective noise contributions. Here, for simplicity, and to clarify the main results of our findings, we assume that $\tau$ and $\rm{f}$ are constant parameters and restrict the investigation of noise contributions to the global offset parameter, i.e. $y_0(t)$, which we assume to be normally distributed [$\left\langle y_0(t)\right\rangle = 0$, $\left\langle y_0^2(t)\right\rangle = \sigma_{y_0}^2$]. \\
\indent In Fig\,\ref{fig:Example Ringdown} we show an example of a discretely sampled damped sinusoidal signal. For the analysis of such a signal we consider four key parameters that affect the expected precision and accuracy: a) the number of signal oscillations per typical decay time, $\rm{f}\times \tau$; b) the number of samples per typical decay time, $n \times \tau^{-1}$ (i.e. the sampling rate); c) the number of decay times measured in a measurement time window $T_{\text{m}}$, $T_{\text{m}} \times\tau^{-1}$; and d) the signal-to-noise ratio, defined as $\text{SNR} = A \times \sigma_{{y_0}}^{-1}$. 

\subsection{Cram\'er-Rao Lower Bound}
\noindent The fundamental limit for the statistical uncertainty of determining the oscillating frequency of a damped sinusoidal signal (Eq.\,\ref{EQ: FID}) is described by the Cram\'er-Rao lower bound (CRLB)\,\cite{yao1995cramer,gemmel2010ultra}, which sets the lower limit on the variance $\sigma_{\rm{f}}^2$ of any frequency estimator. The CRLB condition for the frequency extracted from a discrete damped sinusoid is given by\,\cite{Yao1995,Gemmel2010,Koch2015,Hunter2018},
\begin{align}
    \sigma^2_{\rm{f}} = \dfrac{6}{(2\pi)^2\,\text{SNR}^2\, {\rm{f}}_{_\text{BW}} \, T_\text{m}^3}\, \chi\left(\tau/T_{\text{m}}\right),
    \label{EQ:CRLB}
\end{align}
where SNR is the signal-to-noise ratio of the signal; f$_{_\text{BW}}$ is the sampling-rate-limited bandwidth of the measurement; $T_{\text{m}}$ is the measurement time window; and $\chi\left(\tau/T_{\text{m}}\right)$ is a correction factor that takes into account the signal decay, which is given by
\begin{equation}
    \chi(r) = \dfrac{e^{2/r} - 1}{3r^3\,\text{cosh}\left(2/r\right)-3r\left(r^2+2\right)}.
    \end{equation}
    
The factor $\chi\left(\tau/T_{\text{m}}\right)$ serves as a compensation factor in Eq.\,\ref{EQ:CRLB} that penalizes measurement of the tails of the exponential decay when the signal has effectively died out. Equation\,\ref{EQ:CRLB} remains valid under the condition that the period of the oscillation is much shorter than the decay time of the signal and that a sufficient number of oscillations occurs in it. Moreover, Eq.\,\ref{EQ:CRLB} dictates that any noise sources affecting the signal detection are contributing to the fundamental CRLB limit through their effect on the SNR of signal. \\
\indent In Ref.\,\cite{visschers2020continuous}, the authors demonstrate that the CRLB limit is the appropriate estimator of the fundamental sensitivity of frequency-based measurements within the context of CRDP, as the frequency measurements are directly translated into polarimetric results. However, one needs to carefully investigate whether different signal processing techniques can approach the CRLB, and if yes, under what conditions this is possible. Moreover, considering our motivation is the development of a portable CRDP instrumentation operating with similar principles as recent demonstrations of it\,\cite{Sofikitis2015SDR,visschers2020continuous}, we focus on investigating and comparing different signal processing approaches in terms of their speed and attainable accuracy and precision for damped sinusoidal signals with decay times in the range of $1-10$\,\textmu s and frequencies in the range of $1-10$\,MHz.

\subsection{Signal Analysis}
\subsubsection{Least-Squares Estimation of Nonlinear Parameters}
\noindent For time-domain analysis we focus on an optimized least squares curve fitting approach based on the Levenberg-Marquardt algorithm (LMA)\,\cite{more1978levenberg}. The algorithm minimizes the sum of the squared residuals,
\begin{align}
    S = \sum_{i=1}^{n} \left[y_{i} -f\left(t_i,\beta\right)\right]^2,
\end{align}
where $y_{i}$ is the $i^{th}$ sample of the discretized recorded signal $y(t)$ (Eq.\,\ref{EQ: FID}), $t_{i}$ is the $i^{th}$ time sample, and $f(x,\beta)$ is the non-linear fit function (Eq.\,\ref{EQ: FID}) with $\beta$ representing the guess fitting parameters for \{$A,\tau, \rm{f}, \phi,\ y_0$\}. The LMA algorithm iteratively finds the optimal guess parameters $\beta$ describing the recorded signal $y$.\\
\indent The LMA is, in itself, an efficient algorithm, but it relies heavily on the initial guess parameters of the iterative process. However, we wish to identify the precision and speed limitations of computational implementations of a least-squares algorithm and, hence, we assume for our computational investigations that the initial conditions are well-defined and known in advance (with our experimental investigation we examine the dependence of the LMA algorithm on the initial guess parameters under realistic conditions; see Sec.\,\ref{sec:Exp}). Furthermore, the time required for the convergence of a fit using LMA is highly dependent on the platform used. In this work we choose to work with a CPU-based code for the implementation of the LMA that employs a Python optimized package (SciPy) based on the MINPACK library\,\cite{More1980}.

\subsubsection{Fast Fourier transform}
\noindent For frequency-domain analysis we use a fast Fourier transform (FFT) algorithm, as introduced by Cooley and Tukey\,\cite{cooley1965algorithm}, to calculate the discrete Fourier transform (DFT) of the signal,
\begin{align}
    \mathcal{F}(k) = \sum^{N-1}_{n=0} y_n \,e^{-\frac{\,2\pi i}{N}n\,k} ,\quad k = 0, \ldots, N-1,
\end{align}
where $y_n$ is the $n^{\text{th}}$ sample of the discretized time-domain signal $y(t)$ (Eq.\,\ref{EQ: FID}). We note here that the simplest and most common implementations of the FFT algorithm introduced by Cooley and Tuckey assume that $N$ is a power of two. \\
\indent The Fourier transform of a monochromatic damped sinusoidal signal corresponds to a single Fourier (frequency) component with a spectral width inversely proportional to the signal's decay time. Our aim is to estimate accurately and precisely the central value of this component, rapidly. One approach is to perform a least-squares curve fitting on the resulting FFT spectrum to obtain the central value of the frequency component and its width. However, the accuracy and precision of such process depends strongly on the curve fit-model selected and its initial guess parameters, but, importantly, the speed of such an approach would be at least equal to the overall time required to perform both the FFT and the least-squares fitting. Furthermore, in order for such an approach to be as precise as the direct time-domain analysis approach using, e.g., LMA, one typically employs additional data manipulation techniques (e.g. zero-padding, apodization). \\
\indent Here, we focus on algebraic approaches for the rapid extraction of the central value of the Fourier (frequency) component from the FFT spectrum. One such approach is to determine the center value of this component by considering the three closest neighbouring points to the maximum frequency value (peak): $(k_i,\mathcal{F}(k_i))\equiv(k_i,b_i)$, ($i=1,2,3$), and use a quadratic estimator to find f$_{\text{max}}$ as:
\begin{align}
    \text{f}_{\text{max}} = \dfrac{k_1^2b_1\left(b_3-b_2\right)+k_2^2b_2\left(b_1-b_3\right)+k_3^2b_3\left(b_2-b_1\right)}{2\left[k_1b_1\left(b_3-b_2\right)+k_2b_2\left(b_1-b_3\right)+k_3b_3\left(b_2-b_1\right)\right]}.
\end{align}
It is important to emphasize that the selection of the neighbouring points is crucial for the accuracy (not the speed) of the frequency estimation using such an approach. For high sampling rates, for instance, one can choose - symmetrically, or even asymmetrically - points further away from the closest neighbouring points to the peak, and preferably points lying near to the half-maximum of the Fourier component [this can be easily pre-set in the algorithm if the decay time and the sampling rate are (approximately) known in advance]. Such an algebraic approach on analysing FFT spectra has already been successfully implemented for rapid frequency estimation in CRDP-based experiments (see Ref.\,\cite{stamataki2013monitoring}). By choosing such an approach, we ensure that the computational speed remains as close as possible to the speed required to employ a FFT algorithm. \\
\indent There exist several CPU-based codes available for FFT analysis, but for an appropriate speed comparison between the alternative signal processing methodologies presented in this work, we use a DFT algorithm directly from a Python-based scientific environment (NumPy; we note here that we do not observe in our analysis any differences between different FFT libraries in Python such as SciPy and NumPy).

\subsubsection{Prony}\label{sec:prony}
The Prony method, is a time domain approach originally designed for processing discrete time signals that are superpositions of damped sinusoids. The Prony method is closely related to the Matrix Pencil method (both estimate the signal as a sum of complex exponentials)\,\cite{Hua1990,Sarrazin2011}, the latter being used in NMR analysis\,\cite{Lin1997,Fricke2020}. However, Prony analysis takes a polynomial approach in parameter (frequencies and damping factors) estimation whereas Matrix Pencil Method locates the signal parameters by finding the eigenvalues to a matrix pencil. \\
\indent The application of the Prony method follows in three steps: (a) an autoregressive model is built employing discrete measurements; (b) the roots of the characteristic polynomial for the corresponding finite difference equation are statistically estimated; and (c) estimates of the parameters of the signal are derived from the roots.\\
\indent For the special case of one samped sinusoid, a discrete time sampling of such a signal gives rise to an autoregressive model of order $3$  where the measurement $y_k$ at time $k$ is expressed through $3$ preceding  measurements in a linear way:
\begin{equation}
y_{k+3}+ \alpha_2 y_{k+2}+\alpha_1 y_{k+1}+ \alpha_0 y_k =0,
\end{equation}
where $k$ varies from $0$ to $n+2$, and $n+5$ is the number of measurements (sampling points). The coefficients $\alpha_i$ are determined by any of the linear systems
\begin{equation}
\begin{bmatrix}
y_{k+2} & y_{k+1} & y_k   \\
y_{k+3} & y_{k+2} & y_{k+1} \\
y_{k+4} & y_{k+3} & y_{k+2}
\end{bmatrix}
\cdot
\begin{bmatrix}
\alpha_2 \\ \alpha_1 \\ \alpha_{0}
\end{bmatrix}
=
-\begin{bmatrix}
y_{k+3} \\ y_{k+4} \\ y_{k+5}
\end{bmatrix}
\label{equ:mat}
\end{equation}
with $k=0,\ldots,n$.\\
\indent In the presence of noise, the $3\times 3$-matrix in the left-hand side and the 3-vector in the right-hand side are random, so the coefficients $\alpha_i$ can be found, e.g., by the least square method minimising the loss function
$$\sum_{k=0}^n(y_{k+3}+\alpha_2y_{k+2}+\alpha_1y_{k+1}+\alpha_0y_{k})^2.$$
These constitute the characteristic polynomial equation
\begin{equation}
q(z) = z^3 + \alpha_{2}z^{2} +\alpha_{1}z  + \alpha_0,
\end{equation}
whose roots $u_0, u_\pm$ incorporate the parameters of the signal. In the special case under study,
$u_0= e^{-\frac{\Delta}{\tau} }$, $u_\pm= e^{-\frac{\Delta}{\tau} } e^{\pm  i 2 \pi \rm{f} \Delta}$,
where $\Delta$ is the sampling time interval. Then the frequency and decay constant of the signal are found as
\begin{eqnarray}
\tau& =& -\ln (u_0)/\Delta,
\\
\text{f} &=& \text{Im}[\ln (u_+)/\Delta].
\label{Prony_estimates}
\end{eqnarray}
The roots can be calculated, e.g., by the Cardano formulas, or by
employing $\alpha_0=-u_0^3$ (in the limit of no noise).
The root $u_+$ can be distinguished from $u_-$ as the one
with positive imaginary part if $u_++u_->0$, and negative otherwise.

A practical realization of this scheme has to take into account the role of the sampling rate $n\times\tau^{-1}=\frac{1}{\Delta}$. Even in the absence of noise, there exist singular values at $n\times\tau^{-1}=\frac{\text{f}}{\pi N}$ (with positive integer $N$)
for which the matrix in Eq.\,\ref{equ:mat} becomes degenerate (degeneracy occurs
for half-integer $N$), and the sampling rate has to be chosen to be different from such singular values.
Another thing is that the coefficients $\alpha_i$ depend on $\text{f}$ via $\cos(\text{f}\Delta)$. Keeping
the sampling rate above $\text{f}_{max}\Delta$ confines $\text{f}\Delta$ in $[0,\pi]$ and
determines $\text{f}$  from Eq.\,\ref{Prony_estimates} uniquely.%

\indent Furthermore, in the presence of noise, the accuracy still depends on  $n\times\tau^{-1}$ even if it exceeds $\frac{\text{f}}{\pi}$. So, if the sampling rate is too high, the sampled points are too close to each other (note that their number is fixed), and small variations of the signal (a smooth function) from point to point are distorted by random jumps which deteriorate estimation. Therefore the frequency should be bounded from below, say with $\text{f}_{min}$. In practice, for a reasonable SNR the dependence of the result on the sampling rate is weak in a wide range of $n\times\tau^{-1}$ values, and this observation can be used for estimation.


\section{Methods}
\subsection{Signal Simulation}
\noindent To compare the three methods of analysis on their respective precision and accuracy in estimating the central frequency of damped sinusoidal signals (Fig.\,\ref{fig:Example Ringdown}), we generate and analyze sets of 500 such signals on a homemade Python CPU-code on a Windows\,10 workstation [CPU: AMD Ryzen 7\,2700, RAM: 16.0\,GB 1330\,MHz DDR4]. All simulated signals have the following non-changing parameter values: $A=1$, $\langle y_0\rangle = 0$, and $\phi = 0$. We also choose the following baseline values for the key parameters of each simulated signal: $\text{f}\times \tau = 5, $\ $n\times \tau^{-1} = 1000/\tau$, $T_{\text{m}}\times \tau^{-1} = 5$, and SNR\,$= 2^{12}$. We choose here a high baseline value for the SNR to clearly examine whether the computational approaches can reach the fundamental CRLB limit as a function of the other key signal parameters. We proceed by varying each key parameter over several orders of magnitude while keeping the other parameters at their baseline value, to explore the dependence of the precision and accuracy of each computational approach on these parameters.

\subsection{Precision and accuracy}
\noindent As a way to quantify the precision of each computational approach we use the standard deviation from the distribution of frequency values obtained through the analysis of the 500 simulated signals, i.e. $\sigma_{\text{f}}$, to estimate the fractional uncertainty $\sigma_{\text{f}}/\rm{f}$ (i.e. smaller fractional uncertainty corresponds to higher precision). Similarly, we define as the accuracy of a method as
\begin{align}
    \text{accuracy} =\frac{1}{\mathcal{N}}\sum _{i=1}^{\mathcal{N}}\dfrac{|\text{f}_{i,\text{est}} - \text{f}_{\text{act}}|}{\text{f}_{\text{act}}},
\end{align} 
where $\text{f}_{i,\text{est}}$ is the frequency estimated by the analysis method for a single signal, $\text{f}_{\text{act}}$ is the actual (input) frequency of the simulated signal (again here, $\mathcal{N}=500$). An analysis method is predicted to have no bias as long as the accuracy of its frequency estimation falls within the precision of the estimation. 

\subsection{Computation time}
We determine the speed of each computation method by estimating the time required to analyze a single signal using the internal timing functions of the Python software (e.g., function \textit{timeit}). 

\begin{figure*}[ht!]
    \centering
    \includegraphics[width=\linewidth]{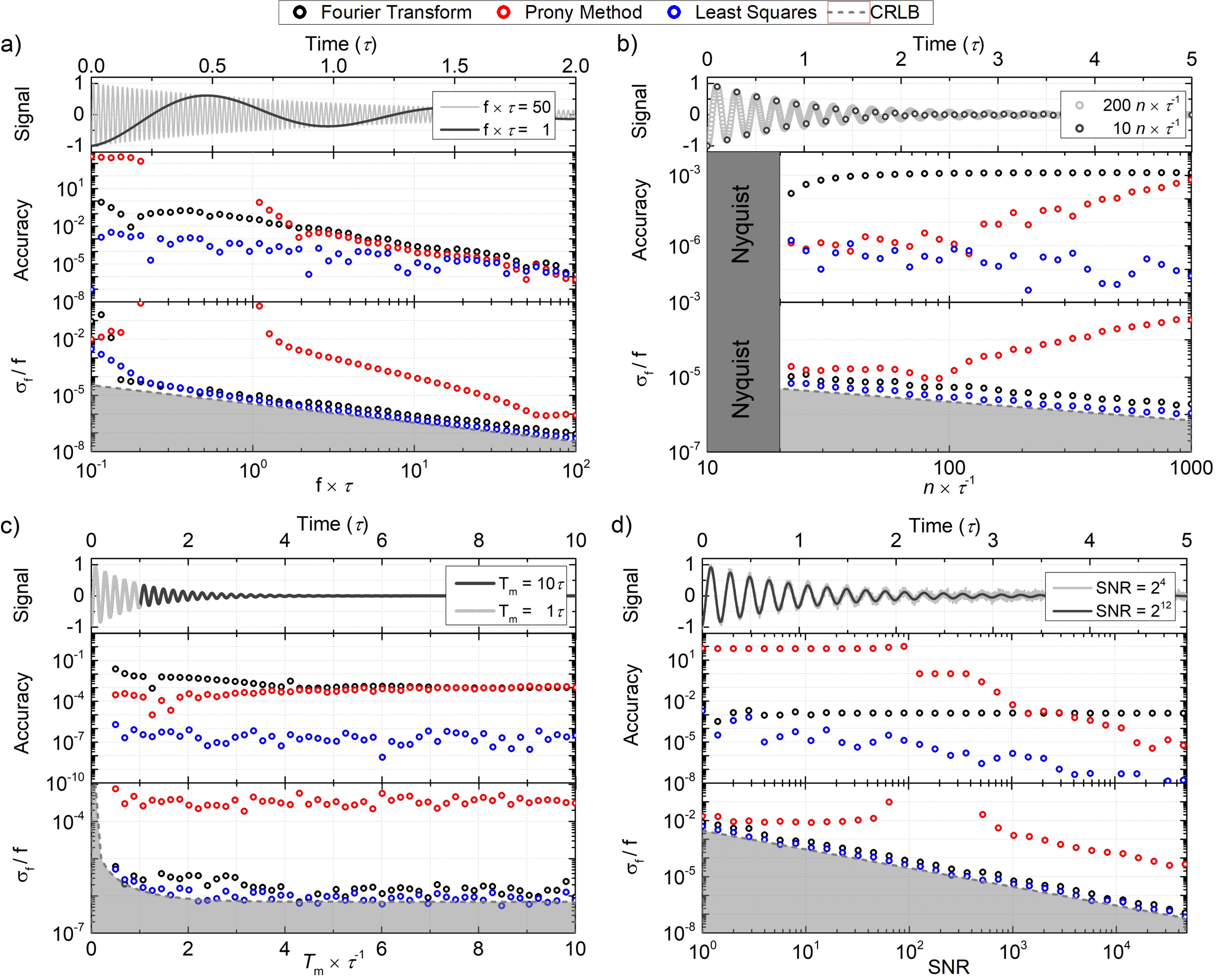}\vspace*{-3mm}
    \caption{\small{Results of attainable fractional uncertainty, $\sigma_{\rm{f}}/\rm{f}$, and accuracy using the FFT (black points), Prony (red points), and least-squares (blue points) analysis methods as a function of varying signal conditions: (a) signal frequency, $\text{f}\times\tau$; (b) sampling rate, $n\times\tau^{-1}$; (c) measurement time-window, $T_m\times\tau^{-1}$; and (d) signal-to-noise-ratio (SNR). In combination with the performance of each method, the fundamental frequency estimation limit given by the Cram\'er Rao lower bound (CRLB) is also shown (dashed gray line). The least-squares method yields results close to the CRLB limit in virtually all situations, with the FFT method yielding similar results, while the Prony yields poor results under most selected conditions.}}
    \label{fig:Precision}
\end{figure*}

\section{Results}
\subsection{Precision and accuracy}
In Fig.\,\ref{fig:Precision} we present results on the precision and accuracy achieved when analyzing discrete simulated damped sinusoidal signals using the least-squares, FFT, and Prony computation analysis methods, as a function of varying signal conditions. In particular:\\
\indent \emph{\textbf{Frequency - }} In Fig.\,\ref{fig:Precision}\,(a) we show a comparison of the attainable precision and accuracy between the three different approaches as a function of the frequency of oscillation. For all frequencies the least-squares approach results in optimal accuracies compared to the other approaches, with the FFT approach being consistently less accurate (this is largely related to the peak-finding algebraic methodology we employ here). The Prony method becomes particularly inaccurate for low frequencies, which is related with the computational formulation of the Prony method that doesn't allows us to investigate a large parameter space without approaching singular points in the analysis. In terms of precision, both the least squares and FFT methods approach closely the CRLB, the latter being approximately a factor of three less precise than the former, while for both methodologies the precision is not influenced by the frequency value (the least squares method deviates from the CRLB at frequencies $\text{f}\times\tau < 0.2$, as expected, since the observed time window does not contain a full period of oscillation). The Prony method yields results with poor accuracy for $\text{f}\times\tau<1$ and does not reach the expected precision limits for the whole simulated frequency range.\\
\indent \emph{\textbf{Sampling Rate -} } In Fig.\,\ref{fig:Precision}\,(b) we show a similar comparison as a function of the sampling rate, i.e. as a function of the number of sample points per decay time $n\times \tau^{-1}$. We note again here that for these estimations we choose a constant frequency of $\text{f}\times\tau^{-1}=5$. As such, the Nyquist criterion limits the lowest sampling rate for a sensible frequency estimate to $n\times\tau^{-1} = 20$. In terms of accuracy and precision the least squares method yields optimal results, while the FFT method yields optimal precision but relatively poor accuracy (at the $10^{-3}$ level), both related to the peak-finding algorithm (these can be improved by performing additional signal manipulation, such as zero padding, but this will significantly affect the computational speed). The Prony method provides accurate and precise results for low sampling rates, but these deteriorate for high sampling rates (see Sec.\,\ref{sec:prony}). We also observe that the least squares method is limited by the CRLB over the entire range of sampling range we investigate, while the FFT method remains consistently less precise.\\
\indent \emph{\textbf{Measurement window -} } In Fig.\,\ref{fig:Precision}\,(c) we present results as a function of the measurement window, i.e. $T_{\text{m}}\times \tau^{-1}$. The least-squares method yields optimal accuracy ($\sim10^{-7}$) and precision ($\sim10^{-5}$) results for $T_{\text{m}}\times \tau^{-1}>2$, however, for short measurement windows the precision becomes poor [as predicted by the CRLB limit, Eq.\,\ref{EQ:CRLB}]. The FFT method reaches its optimum accuracy ($\sim10^{-4}$) and precision for $T_{\text{m}}\times \tau^{-1}\approx4$. Importantly, we observe that both the least-squares and FFT methods reach the CLRB limit for $T_{\text{m}}\times \tau^{-1}>4$, with an optimal measurement window for precise signal analysis using both methods to be $\sim5\tau$. Similar conclusions have already been reported in Ref.\,\cite{everest2008discrete}, suggesting that $\sim5\tau$ can be considered the optimum repetition rate for, e.g., CRDP/CRDE experiments, as compared to the longer acquisition windows ($T_{\text{m}}\times \tau^{-1}>5$) typically required in traditional CRD spectroscopy\,\cite{Huang2013}. The Prony method reaches similar accuracies as the FFT method but its precision is two orders of magnitude larger than the predicted CRLB limit.\\
\indent \emph{\textbf{SNR - }} The final key parameter we vary is the signal's SNR, with the results seen in Fig.\,\ref{fig:Precision}\,(d). The least-squares analysis yields results close to the CRLB limit, while the precision attained using FFT analysis method is approximately a factor of two ($\times2$) higher. Notwithstanding, we see that for an optimum measurement time-window of $5\tau$ [Fig.\,\ref{fig:Precision}\,(c)] and a $\text{SNR}\approx2^{12}$ both the least-squares and FFT methods yield precisions at the $\sim10^{-6}$ levels. In contrast, the Prony method does not provide reliable frequency estimates for signals with SNR\,$<500$, while, for higher SNRs, the attainable precision is two orders of magnitude above the CRLB limit.


\subsection{Speed}
\begin{figure}[h!]
  \centering  \includegraphics[width=0.89\linewidth]{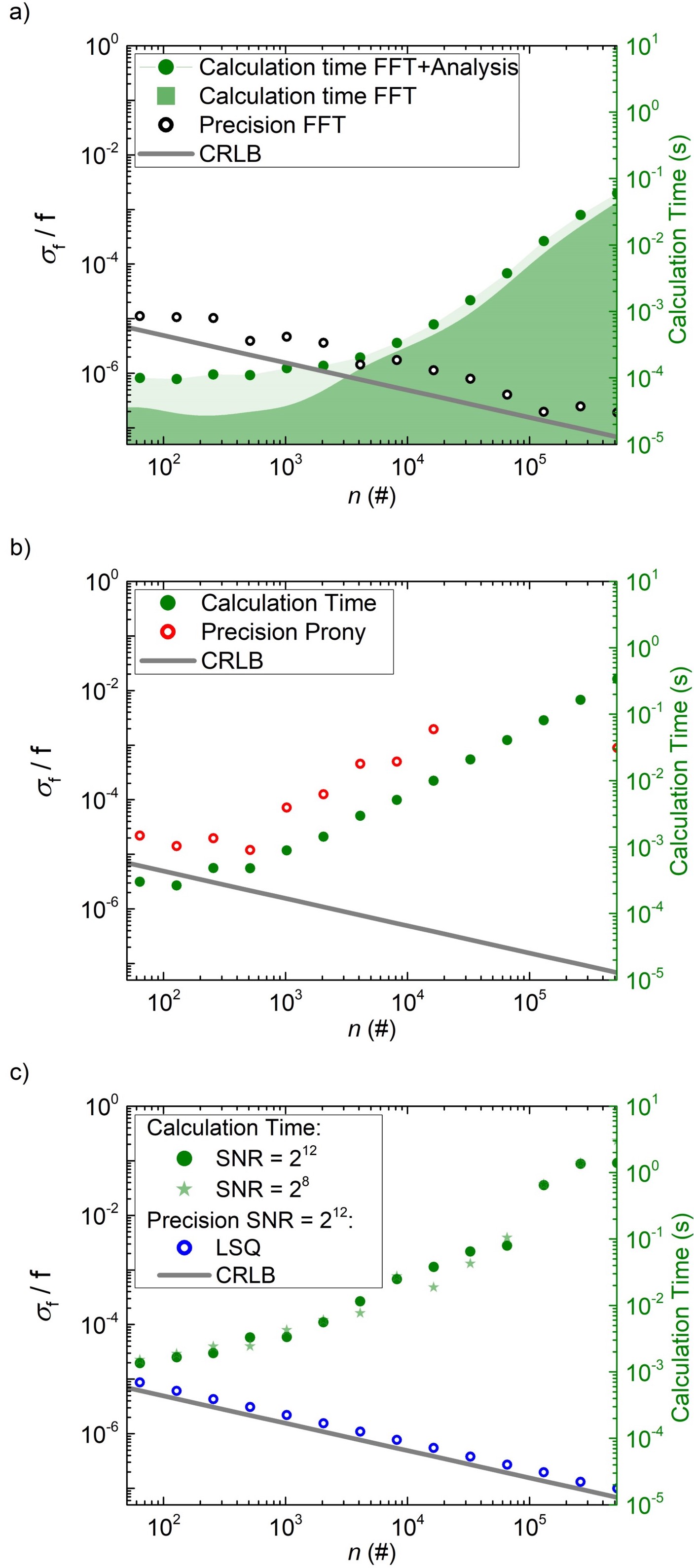}\vspace*{-3mm}
    \caption{\small{Dependence of precision and calculation time (green points) on the number of samples for the: a) FFT (black points), b) Prony (red points), and c) least-squares (blue points) evaluation methods. In all cases, we compare the obtained results with the expected Cram\'er-Rao lower bound limit (CRLB; solid gray line) for increasing sample size [$n\,(\#)$] of simulated signals with $\text{f}\times\tau = 5$, SNR$\,=2^{12}$ and $T_m\times\tau^{-1}=5$. Using the FFT algorithm methodology, for a signal with $10^3$ data points we obtain fractional uncertainties at the sub-$10^{-5}$ levels within a computational time of $\sim200\,\mu$s, while for the same conditions using a least-squares approach we obtain similar precisions but for a computational time of $\sim5$\,ms. In overall, the Prony method results in poor sensitivities and with $>$\,ms computational times.}}
    \label{fig:Speed}
\end{figure}
In Fig.\,\ref{fig:Speed} we present results on the dependence of the fitting (computation) time for each method on the number of data points in a single damped sinusoidal signal, which we also compare with the attainable precision for each case. For these simulations, we use the results presented in Fig.\,\ref{fig:Precision} to choose optimum values for the signal's key parameters: $\text{f}\times\tau = 5$, SNR$\,=2^{12}$ and $T_m\times\tau^{-1}=5$ (such values are also realistically attainable in experiments; see discussions in Ref.\,\cite{visschers2020continuous}).\\
\indent Overall we observe a non-linear increase in the calculation time as the number of samples is increased, with the least-squares and Prony algorithms being more than an order of magnitude slower than the FFT+peak-finding algorithms. In addition, while the least-squares method reaches the CLRB limit, the FFT method yields fractional uncertainties approximately a factor of two larger than the predicted CRLB limit, with the Prony method practically never reaching optimal precision levels. Most importantly, we observe that for a discrete signal with $\sim10^3$ sample points, using the FFT+peak-finding algorithm one can achieve ppm sensitivities ($10^{-6}$ fractional uncertainties) for computational times of $\sim200$\,\textmu s. Under the same conditions, the least-squares algorithm requires approximately $\sim5$\,ms to reach similar fractional uncertainties. In addition, we observe that under similar conditions the Prony method results in poor sensitivities ($10^{-4}$ fractional uncertainties) requiring long ($>$\,ms) computational times.
\begin{figure*}[ht!]
   \centering \includegraphics[width=0.8\linewidth]{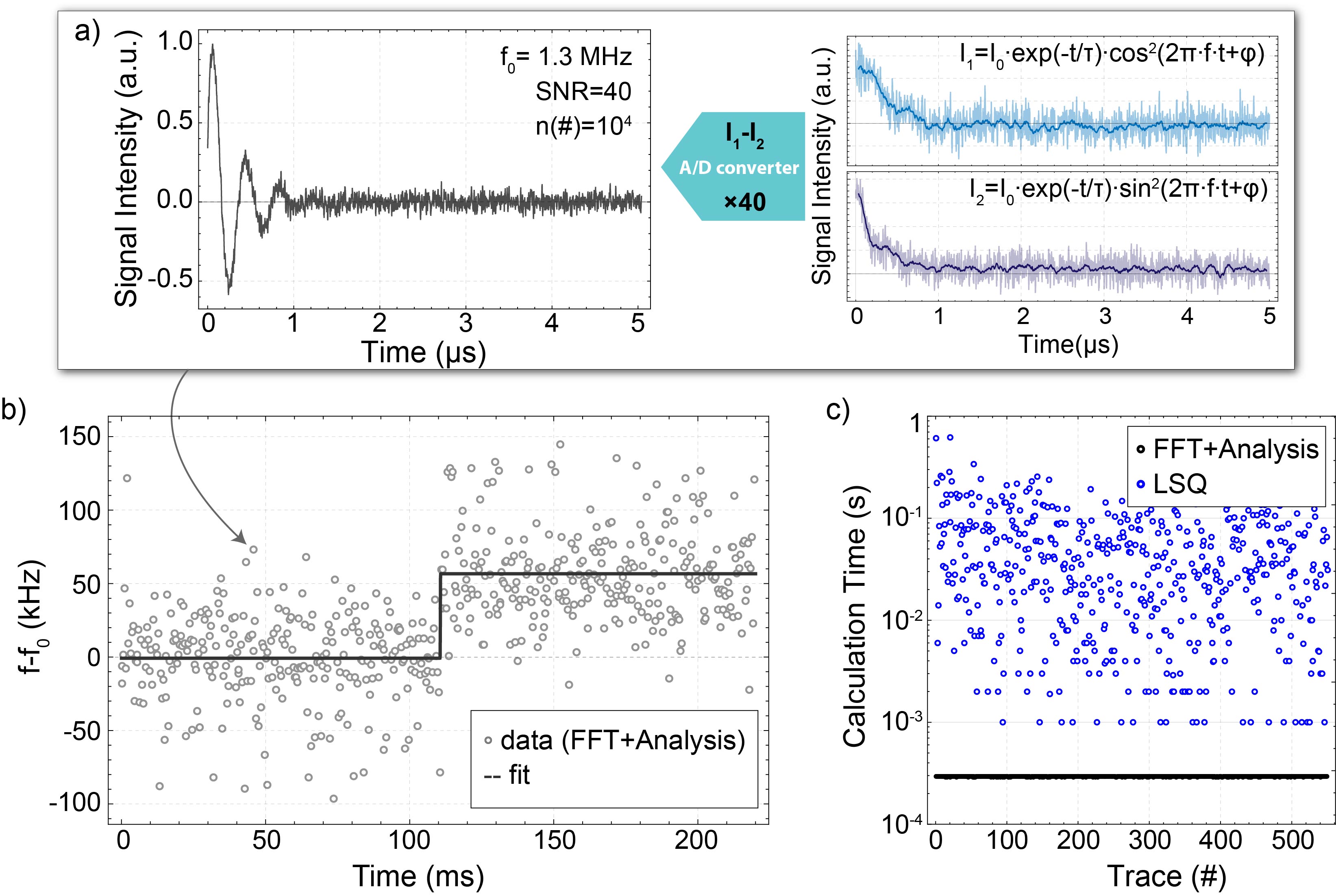}\vspace*{-3mm}
    \caption{\small{\textit{Rapid analysis of experimental signals:} a) CRDP experimental signals showing polarization beat frequencies generated via the Faraday effect on a SiO$_2$ substrate, as recorded by two orthogonal channels of a linear balanced polarimeter (right side). By subtracting these and averaging 40 consecutive traces, at a repetition rate of 100\,kHz, we acquire a CRDP trace (i.e. damped sinusoidal oscillation) with a SNR=40 within 400\,\textmu s. (b) Direct observation of frequency shifts with respect to the (bias) Faraday polarization beat frequency [$\rm{f}_0=1.3$\,MHz], as a function of an externally applied, rapidly pulsed, magnetic field: each frequency (measurement) point is the result of an online FFT analysis of a CRDP trace, as the one shown in (a), and requires $\sim300$\,\textmu s of calculation time [(c)]. The (black) line is the result of a nonlinear least-squares regression analysis used to fit a sigmoid function to the data, demonstrating that we can resolve sub-\textmu s dynamics in a running-fashion. Note that the CRLB estimation for such signals (with $10^4$ data points, SNR=40, and $\tau\approx0.33$\,\textmu s) is $\sim1.3$\,kHz (Eq.\,\ref{EQ:CRLB}), and the observed scattering is associated with experimental noise. (c) Computational time required to perform the FFT+peak-finding analysis algorithm is $\sim300$\,\textmu s per trace and remains constant during the analysis of the complete set of traces, while a least-squares approach requires computational times ranging from $\sim1$\,ms$\,-\,$1\,s (since least-square fitting is highly dependent on the initial fit parameters).}} 
    \label{fig:Experiment}
\end{figure*}

\subsection{Experiment}\label{sec:Exp}
As a proof-of-principle demonstration for the capabilities of our analysis methodology we use a CRD-based polarimetric instrument we have developed in our laboratory for the attainment of experimental CRDP, i.e. damped sinusoidal, signals (see for details Ref.\,\cite{visschers2020continuous}).\\
\indent Briefly, the ring-down cavity of the instrument has a total length of 0.60\,m and consists of two concave mirrors with radii of curvature of 1\,m and specified reflectivity R$\sim$99.9\% at 408\,nm (FiveNine Optics). We use a single-frequency CW laser source (Toptica DL-PRO; $\lambda = 408$\,nm) that we rapidly pulse to initiate ring-down events [with the use of an acousto-optic modulator (AOM; Gooch and Housego 3200-125)]. In our optical setup we can generate CRDP signals with ring-down times in the 0.3-1.5\,\textmu s range (depending on the usage of intracavity optics), at repetition rates as high as 100\,kHz, that we record and digitize using a 14-bit digitizer (Teledyne, ADQ14DC-2X-PCIE, dual channel DC-coupled operation; sample rates of 2\,GS/s per channel), which has a maximum acquisition rate of 100\,kHz (mainly limited by the data transfer rate), 14-bit resolution per channel, and permits on-board channel subtraction and signal averaging. \\
\indent For our demonstration, we use the (non-resonant) Faraday effect of a 6.35(1)\,mm thick, AR-coated SiO$_2$ substrate (FiveNines Optics; AR coated by FiveNine Optics with specified  R$<0.01$\%). In particular, using permanent magnets directly attached to the substrate, we can generate large enough Faraday optical rotations $\thetaF$\,\cite{bougas2012,bougas2015chiral,visschers2020continuous} that result in CRDP signals with polarization beat frequencies in the range of 1-3\,MHz [the beat frequency is proportional to the induced Faraday rotation as: $\rm{f}=\theta_{\rm{F}}\cdot\rm{FSR}/\pi$, where FSR$=(c/L_{rt})$ is the cavity's free spectral range, with $c$ the speed of light and $L_{rt}$ the round-trip cavity length].\\
\indent To demonstrate our ability to monitor and analyze CRDP experimental signals in a running fashion at $\sim$kHz rates using the FFT+peak-finding analysis algorithm [in accord with the results shown in Fig.\,\ref{fig:Speed}\,(a)], we proceed as follows: we initiate ring-down events at a rate of 100\,kHz and continuously record the polarimetric ring-down signals (i.e., the photo-detector signals) for $\approx220$\,ms, while channel subtraction allows for the generation of damped sinusoidal signals (CRDP traces) and on-board signal averaging enables the average of 40 consecutive traces; each trace, therefore, requires an integration time of 400\,\textmu s. In Fig.\,\ref{fig:Experiment}\,(a) we show such an experimentally acquired CRDP trace with a Faraday-rotation-related polarization beat frequency of $\rm{f}_0=1.3$\,MHz, a ring-down time of $\tau=0.324(2)$\,\textmu s, and an SNR\,$\simeq40$. Note that, given the digitizer's sampling rate, the CRLB limit in estimating the signal's beat frequency is 1.3\,kHz, i.e. $\sigma_{\rm{f}}/\rm{f}=10^{-3}$ (Eq.\,\ref{EQ:CRLB}; Fig.\,\ref{fig:Precision}). Furthermore, using an additional, homemade, solenoid [with a length of $2.53(1)$\,cm, and a diameter of $3.05(1)$\,cm] placed around the SiO$_2$ substrate, we induce a rapid frequency shift on the recorded CRDP signal by applying a (rapidly pulsed) external magnetic field [using a USB controlled metal-oxide-semiconductor field-effect transistor-based switching circuit resulting in switch-on times of $<10$\,\textmu s]. \\
\indent In Fig.\,\ref{fig:Experiment}\,(b)\,\&\,(c) we show our experimental results. We see that using an online FFT+peak-finding approach we can analyze CRDP traces at a constant rate of $\sim$3.3\,kHz, which is comparable to the acquisition rate of the individual CRDP traces, and we can sensitively follow parameter changes - in our case, frequency changes - at \textmu s time scales [evident from the analysis of our results using a sigmoid function that yields a $\approx10$\,\textmu s rise-time; Fig.\,\ref{fig:Experiment}\,(b)]. We emphasize here that the frequency fluctuations present in the recorded signals [Fig.\,\ref{fig:Experiment}\,(b)] are the result of experimental noise sources (see Ref.\,\cite{visschers2020continuous}). As a comparison, in Fig.\,\ref{fig:Experiment}\,(c) we also demonstrate that a least-squares approach would require computational times ranging from $\sim1$\,ms to 1\,s to analyze the same CRDP traces and, hence, we would be unable to analyze such a stream of traces online in a running fashion (this wide range of calculation times is related to the dependence of the least-square fitting algorithm to the initial guess fit parameters).

\section{Discussion \& Conclusion}
In this work, we consider different time- and frequency-domain-based computational algorithms for the rapid estimation of the signal parameters of damped sinusoidal signals. We analyze their accuracy and precision in terms of key signal parameters and estimate the computational time required to obtain these using standard computational platforms (e.g. a desktop computer) and software (e.g. Python). Overall, we see that a time-domain-based least-squares algorithm reaches the expected fundamental estimator limits in terms of precision and accuracy, and requires ms-long computational times to obtain fractional uncertainties at the ppm levels (for signals with high SNR), while a Fourier-based algorithm can achieve similar sensitivities at, at least, an order of magnitude faster computational times, even when one employees \textit{standard} computational platforms and algorithms. We also consider in comparison to the least-squares and FFT analysis methods an alternative computational approach based on the Prony method, recently proposed for rapid analysis of damped sinusoidal signals; we observe that an implementation of the Prony method for our parameter-range of interest, fails to provide comparable accuracies and precisions to the least-squares and FFT methods, particularly within similar computational time-scales. We then validate our results using an experimental CRDP setup and a FRGA-based acquisition system, to demonstrate the online recording and analysis of damped sinusoidal signals in a running fashion at $\sim$\,kHz rates using an FFT+peak-finding alogrithm, and, in particular, we demonstrate the ability to observe signal changes at time scales as fast as 10\,\textmu s. \\
\indent Overall, based on our results, we recommend adopting FFT analysis approaches for rapid parameter analysis of damped sinusoidal signals within the context of CRDP/CRDE (and similar) techniques, which offers an optimum combination of speed and performance.\\
\indent As a concluding remark we note that the exact computational speeds for the presented analysis methods depend significantly on the computational platform and software used. In Ref.\,\cite{wilson2019implementation}, the authors demonstrate more than an order of magnitude improvements in computational times for the FFT and Prony methods by implementing alternative computational packages/softwares. As such, we expect that an optimized FFT algorithm may very well outperform our results by more than an order of magnitude, suggesting, therefore, that even with the use of standard data-acquisition systems, highly precise microsecond-resolved signal parameter estimation is possible. This becomes particularly important for our application of interest, CRDP: using the results presented in Ref.\,\cite{visschers2020continuous} we can estimate that in CRDP experiments with typical decay times constants of a few \textmu s and oscillating frequencies of $>$1MHz (typically corresponding to polarimetric signals of a few rad; Refs.\,\cite{sofikitis2014evanescent,bougas2015chiral,visschers2020continuous}), and which can yield (acquired) signals with SNR$\,>2^{12}$, we see that it is possible to perform online, in a running-fashion, microsecond-resolved experiments with \textmu rad polarimetric sensitivities (i.e. sub-nrad/$\sqrt{\rm{Hz}}$ sensitivities). Such a possibility is of paramount importance for emerging applications in chiral sensing and analysis, in surface catalysis, and indicates that real-time monitoring of gas/liquid flows in GC/HPLC and of surface (chiral) dynamics is nowadays feasible. Finally, we anticipate that an overall improvement in computational speeds could also result from implementing a network theory approach for rapid (spectroscopic/spectropolarimetric) signal analysis; for intance, network theory has been recently used for the classification and search of spectral features of various molecules\,\cite{Zaleski2018,Tobias2020}. However, rapid parameter extraction of time-sampled signals could prove to be nontrivial with a network approach, and we will investigate such a possibility in future works.\\


\section*{Acknowledgments}
\indent This work was supported by the European Commission Horizon 2020, project ULTRACHIRAL (Grant No. FETOPEN-737071), and by the European Union's Seventh Framework Programme for Research, Technological Development, and Demonstration, under the project ERA.Net RUS Plus, grant agreement no.\,189 (EPOCHSE). EW, TC, and AM were supported in part by Innovate UK\,KTP\,010819. LB and JV are grateful to Dmitry Budker for his constant help, support and comments. LB is grateful to Michael Everest for his help and support, and to Amelia Meath for fruitful discussions.

\bibliography{LBFastAnalysisCRDP}
\end{document}